\documentclass[
    ,final            
  ]
  {aipproc}

\layoutstyle{6x9}

\begin{document}

\newcommand{\slow}{$\sqrt{s}$ = 200~GeV }
\newcommand{\shigh}{$\sqrt{s}$ = 500~GeV }

\title{The PHENIX Muon Trigger Upgrade}

\classification{13.88+e, 14.70.Fm, 12.15.-y}
\keywords      {RHIC, PHENIX, polarized protons, W boson, spin}

\author{John Lajoie for the PHENIX Collaboration}{
  address={Iowa State University, Department of Physics and Astronomy, Ames, Iowa 50021}
}

\begin{abstract}
The PHENIX muon trigger upgrade adds Level-1 trigger detectors to existing forward 
muon spectrometers and will enhance the ability of the experiment to pursue a rich program of spin physics in polarized proton 
collisions at \shigh. The additional muon trigger detectors and Level-1 trigger electronics will allow the experiment to select 
high momentum muons from the decay of $W$-bosons and reject both beam-associated and low-momentum collision background, enabling the study of 
quark and antiquark polarization in the proton. The muon trigger upgrade will add momentum and timing information to the present muon Level-1 trigger, 
which only makes use of tracking in the PHENIX muon identifier (MuID) panels. Signals from three new resistive plate chambers (RPC's) 
and re-instrumented planes in the existing muon tracking (MuTr) chambers will provide momentum and timing information for the new Level-1 trigger. 
An RPC timing resolution of $\sim$2ns will permit rejection of beam related backgrounds. 


\end{abstract}

\maketitle


\section{Physics with Polarized W Bosons}

A central goal of high-energy physics is to understand the quark and gluon structure of QCD bound states. The most fundamental of 
these bound states is the nucleon, and current measurements indicate that only about 25\% of the spin of this object is 
carried by its quark content \cite{Lampe2000}. Contributions from gluons, orbital angular momentum, and the sea quarks are 
poorly understood. Measurements of the single longitudinal spin asymmetry $A_{L}$ in $W$-boson production
in polarized proton collisions will make it possible to better understand the contributions of the of the sea and valence quarks to 
the spin of the proton. 

\section{The PHENIX Experiment}

The PHENIX experiment is one of two large experiments at the Relativistic Heavy Ion Collider (RHIC) at Brookhaven National 
Laboratory. The experiment consists of two central arms designed for measuring leptons, photons, and 
charged hadrons produced at midrapidty as well as two muon arms at forward and backward rapidities. The muon arms themselves consist of a set of 
tracking chambers (the Muon Tracker, or MuTr) in an approximately radial magnetic field followed by a Muon Identifier (MuID) which consists of layers of absorber 
interspersed with Iarocci tubes.  

RHIC is an extremely versatile accelerator complex, capable of accelerating Au and lighter nuclei as well as polarized protons with $\sim$60\% 
 polarization \cite{Bai}. The PHENIX experiment has an active spin physics program, including measurements of the spin structure of the nucleon
through $\Delta G$, $\Delta q/q$, $\Delta \overline{q}/\overline{q}$ and transversity measurements $\delta q$ \cite{Barish}. While measurements of $\Delta G$ have already
been reported \cite{PHENIX_DG} based on running at \slow, measurements of the quark and antiquark distribution functions will be made in 
future running at \shigh using polarized $W$-bosons, identified through their decay into high momentum muons. The PHENIX experiment has
excellent data acquisition (DAQ) and trigger capabilities, and is capable of a sustained Level-1 accept rate of 5kHz with very low deadtime. This 
combination of a selective trigger and high bandwidth allows the PHENIX experiment to take data for multiple physics signals in parallel. 

\begin{figure}
  \includegraphics[height=.25\textheight]{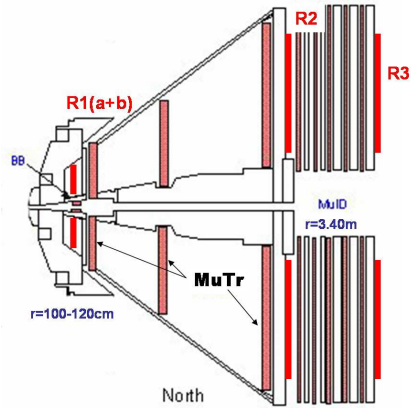}
  \caption{Diagram of a PHENIX muon arm, showing the existing magnet, MuTr and MuID detectors as well as the location of the three RPC chambers to be installed.}
  \label{muon_arm}
\end{figure}

\section{The Muon Trigger Upgrade}

In future running at \shigh it is expected that luminosities near $2 \times 10^{32} cm^{-2} s^{-1}$ will be achieved, corresponding to an interaction rate of
$\sim$12MHz. This high event rate, coupled with the requirement that muon triggers consume no more than 2kHz of the available DAQ bandwidth, will require
an event rejection $>10000$ at Level-1. While PHENIX currently has an existing Level-1 muon trigger based only on the MuID, the achieved rejection at \slow of 
$\sim$250-500 is inadequate. In addition, the rejection for the MuID based trigger is highly sensitive to beam-related background processes in the collider. 

While there are many sources of low-momentum muons in proton-proton collisions (mainly charm and beauty decays), above a transverse momentum of 20~GeV/c the 
muon spectrum at \shigh is dominated by decays of $W$ bosons. In order to achieve the desired event rejection momentum selectivity at Level-1 is required. 
The PHENIX collaboration plans to add this momentum selectivity through a combination of additional instrumentation in the existing PHENIX muon arms.
First, three resistive plate chamber (RPC) tracking chambers will be installed in the PHENIX muon arms, as shown in Figure \ref{muon_arm}. 
All three RPC chambers will use strip readout, with the strips organized as 360 segments in azimuthal angle around 
the beam axis, and between four and six segments in polar angle theta. (Only two theta segments will be used in the trigger.) This portion of the 
upgrade is funded by a grant from the U.S. National Science Foundation.

While RPC development and testing have been ongoing at test stands at the University of Illinois, the University of Colorado, and Georgia State University, we
plan to make use of existing RPC designs from the CMS experiment at the Large Hadron Collider. By doing this we 
leverage the considerable man-years of research and development experience obtained by the CMS collaboration and accelerate the schedule for 
deploying the detectors in PHENIX.  

\begin{figure}
  \includegraphics[height=.17\textheight]{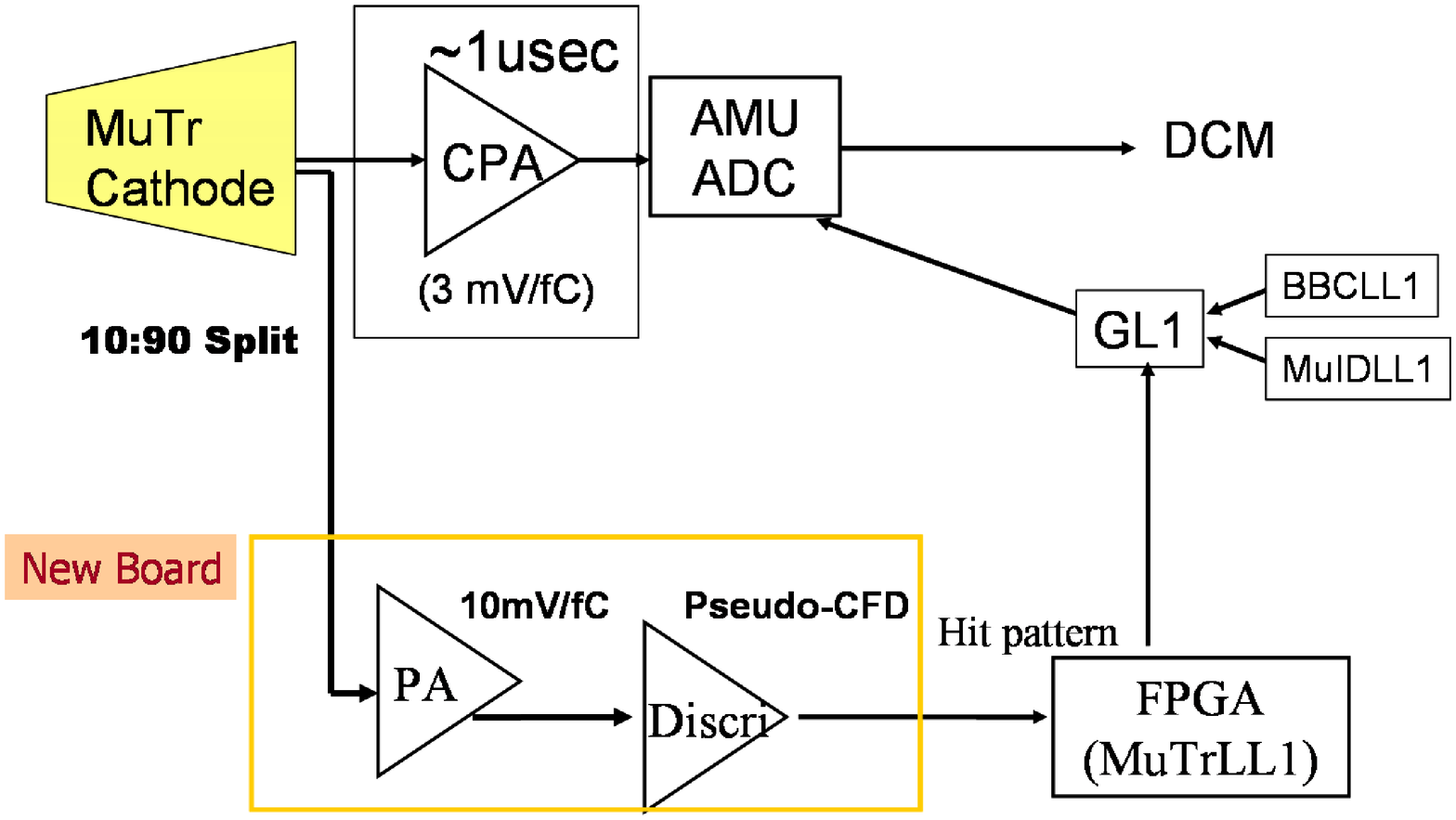}
  \caption{Schematic view of the signal split from the existing MuTr station 2 cathode plane.} 
  \label{mutr_split}
\end{figure}

In addition to the additional information from the RPC chambers, the PHENIX muon trigger upgrade will also include the ability to make use
of high-resolution tracking from the existing MuTr stations by splitting the signal from the chambers and adding an additional 
electronics chain to provide information to the Level-1 trigger (see Figure \ref{mutr_split}). Ongoing tests with a MuTr chamber 
at Kyoto University have demonstrated
that this split can be achieved with passive electronics in such a way that it does not significantly degrade the resolution of the 
MuTr cathode planes.  This portion of the upgrade is funded by the JSPS. 

Momentum selectivity in the PHENIX muon trigger upgrade is achieved by matching hits in the first and second RPC stations and making
a straight line projection into MuTr station 2. Because of the radial magnetic field in the PHENIX muon arms, tracks will be bent 
along the measurement direction in azimuthal angle, so that the deviation of a hit found in MuTr station 2 and the straight line projection
is an indication of the momentum of the track. A cut on all candidates such that the MuTr station 2 hit deviates from the projection by less 
than three cathode strips is efficient for momenta above 20 GeV/c and has been demonstrated to achieve event rejection factors $>10000$ in simulations
using the pythia \cite{pythia} event generator.  It should be noted that in addition to a candidate track in the muon trigger upgrade, the
existing MuID-based muon trigger is also required.  

\begin{figure}
  \includegraphics[height=.2\textheight]{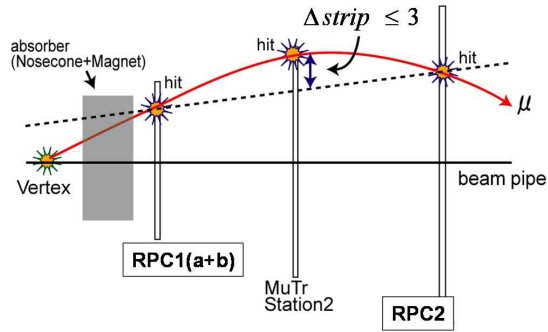}
  \caption{The PHENIX muon trigger upgrade algorithm. Hits in RPC 1 and RPC 2 are used to project into MuTr station 2 and matched to hit.}
  \label{trig_alg}
\end{figure}

As was mentioned previously, at RHIC substantial beam-related collider backgrounds degrade the performance of the existing MuID based trigger. Because the 
magnitude of the background is unknown at \shigh running at RHIC, it is essential that the trigger be designed to be insensitive to such 
backgrounds. In order to achieve this, we will take advantage of the good timing resolution of the RPC chambers, in particular RPC 3, to make a timing 
cut on the RPC hits such that only hits consistent with tracks originating from the vertex are considered in the trigger. This will eliminate 
hits (and backgrounds) from particles that arrive coincident with the beam bunches in the collider.  

\section{Outlook}

\begin{figure}
  \includegraphics[height=.2\textheight]{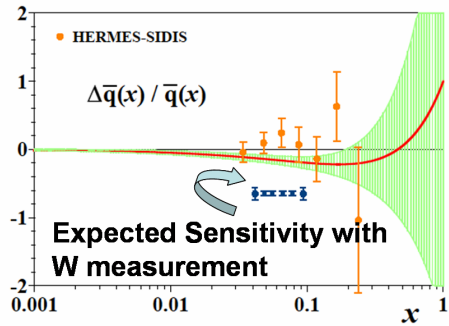}
  \caption{Expected sensitivity of the RHIC measurements using polarized $W$-bosons, assuming the full goals of the RHIC \shigh program. The theory curve
and uncertainty band are from an AAC analysis of existing data.}
  \label{expected}
\end{figure}

It is expected that the current RHIC polarized proton program at \slow will be completed in 2009, with the \shigh program beginning at this point and
continuing through 2012. Luminosity is expected to improve througout the \shigh program, with a total expected integrated luminosity 950 $pb^{-1}$ and 
beam polarization at 70\%. The PHENIX muon trigger upgrade will be in place for the start of high energy polarized proton running in 2009. 
Expected sensitivity for the polarized $W$-boson program at RHIC for $\Delta \overline{q}/\overline{q}$ are shown in Figure \ref{expected}.



\end{document}